\documentclass[aps,prl,reprint,preprintnumbers,amsmath,amssymb,floatfix]{revtex4-1}
\usepackage{longtable}
\usepackage{morefloats}
\usepackage[dvips]{graphicx}
\usepackage{color}
\usepackage{amsmath}
\usepackage{amssymb}
\usepackage{amsfonts}
\usepackage{epsfig,graphicx,amsfonts,amsbsy}
\usepackage{amsmath,amsfonts,amsthm,amssymb}
\usepackage{appendix}
\usepackage{bbm}
\usepackage{amsmath}
\usepackage[bb=dsserif]{mathalpha}
\usepackage{bm}

\usepackage{makeidx}
\usepackage{url}
\usepackage{mathrsfs}  
\usepackage{verbatim}
\usepackage[bookmarksnumbered,pdfpagelabels=true,plainpages=false,colorlinks=true,linkcolor=blue,citecolor=blue,urlcolor=blue]{hyperref}
\usepackage[rightcaption]{sidecap}
\usepackage{array}
\usepackage{booktabs}
\usepackage{multirow}
\usepackage{bbm}
\usepackage{tabularx}
\usepackage{cancel,soul,ulem}

\newcommand{\vv}[1]{\boldsymbol{#1}}

\graphicspath{{figures/}}

\makeatletter

\begin{document}
\def \cedenna {Centro  de Nanociencia y Nanotecnología CEDENNA, Avda. Ecuador 3493, Santiago, Chile}
\def \fcfm {Departamento de F\'isica, FCFM, Universidad de Chile, Santiago, Chile.}
\def \usach {Departamento de F\'isica, Universidad de Santiago de Chile.}
\def \uta {Departamento de Física, Facultad de Ciencias, Universidad de Tarapacá, Casilla 7-D, Arica, Chile}

\author{Carlos Saji${}^{1}$}
\author{Eduardo Saavedra${}^{2}$}
\author{Vagson L. Carvalho-Santos${}^{3}$}
\author{Alvaro S. Nunez${}^{1}$}
\author{Roberto E. Troncoso${}^{4}$}

\affiliation{${}^{1}$\fcfm}
\affiliation{${}^{2}$\usach}
\affiliation{${}^{3}$Departamento de F\'isica, Universidade Federal de Vi\c cosa, 36570-900, Vi\c cosa, Brazil}
\affiliation{${}^{4}$\uta}

\date{\today}
\title{Static and Dynamics of Twisted Skyrmion Tubes in Frustrated Magnets}
\begin{abstract}
Stable three-dimensional topological skyrmion structures in frustrated magnets are investigated. The texture exhibits a helicoid pattern along the vertical direction, described by a position-dependent helicity, which interpolates between Neel- and hedgehog-like two-dimensional skyrmions, characterized by the Hopf index, and is referred to as "twisted skyrmion tubes" (TSkTs). The stability and topology of TSkTs are achieved by competing next-nearest-neighbor exchange interactions, the thickness of the magnet, and the applied magnetic field.
The dynamical behavior of a twisted structure in frustrated magnets is determined. Specifically, we derive that the helicity dynamics of the TSkT can be driven by an electric current resulting from spin-orbit torque interaction. Furthermore, we address the study of the electronic scattering problem using a spin-orbit-torque-driven TSKT, which offers promising applications for low-power storage nanodevices and nanobatteries with enhanced control.
\end{abstract}

\maketitle
{\it Introduction.--} Three-dimensional (3D) nanomagnetism is a cornerstone in modern magnetism research, offering new degrees of freedom for controlling spin textures and magnetic interactions \cite{Pacheco2017,Makarov2022,Pacheco2025}. This research area explores the behavior of 3D magnetic textures, which consist of complex space-dependent arrangements of magnetic moments that extend across all spatial dimensions in a material \cite{Gbel2021} and couple to the geometry of the magnetic body, mirroring its curvature and topology \cite{Hernandez2020,Bezsmertha2024}. Unlike two-dimensional structures such as magnetic bimerons \cite{Tretiakov-PRB-2007,Zhang-PRB-2020,Castro2025}, helical states \cite{Mella2024} or skyrmions \cite{Nagaosa2013, RoldnMolina2016, Troncoso2014a, Troncoso2014b, JaeschkeUbiergo2019, CastilloSeplveda2019} confined to thin films \cite{Nagaosa2013}, 3D textures such as hopfions \cite{Sutcliffe,Zheng2023, Saji2023}, Bloch points \cite{Dring1968, ZambranoRabanal2023, CarvalhoSantos2015, Elias2014, Tapia2024}, magnetic bobbers \cite{Rybakov2015}, or knotted skyrmions \cite{Sutcliffe2017} exhibit richer topologies and can possess unique physical properties due to their geometry and topology \cite{Azhar2004}. These structures are of significant interest in spintronics, specifically in magnetic storage technologies, for their potential in robust, energy-efficient data manipulation.

Magnetic skyrmions have emerged as a compelling topic in condensed matter physics and spintronics due to their unique topological properties, nanoscale size, and potential applications in information and communication technologies \cite{Nagaosa2013}. Skyrmions are topologically protected spin configurations that exhibit a vortex-like arrangement of magnetic moments, stabilized by chiral interactions such as the Dzyaloshinskii-Moriya interaction (DMI) in non-centrosymmetric materials \cite{Fert2017}. Their 3D counterparts, known as skyrmion tubes (or skyrmion strings) \cite{Xing2020}, can be visualized as a skyrmion extended along the third spatial dimension, forming a filament-like structure through the magnetic medium. Skyrmion tubes exhibit rich low-energy translational dynamics \cite{Kravchuk}, have been experimentally observed in chiral magnets \cite{Yu2024}, and are predicted to play a significant role in the dynamics and interactions within skyrmion lattices \cite{Birch2020,Xia2021,Jiang2024,Fullerton2025}. Understanding the formation, stability, and manipulation of Skyrmion tubes is essential for the development of 3D spintronic architectures and for realizing complex functionalities in magnetic devices.

\begin{figure}[!htb]
  \centering
\includegraphics[width=\linewidth]{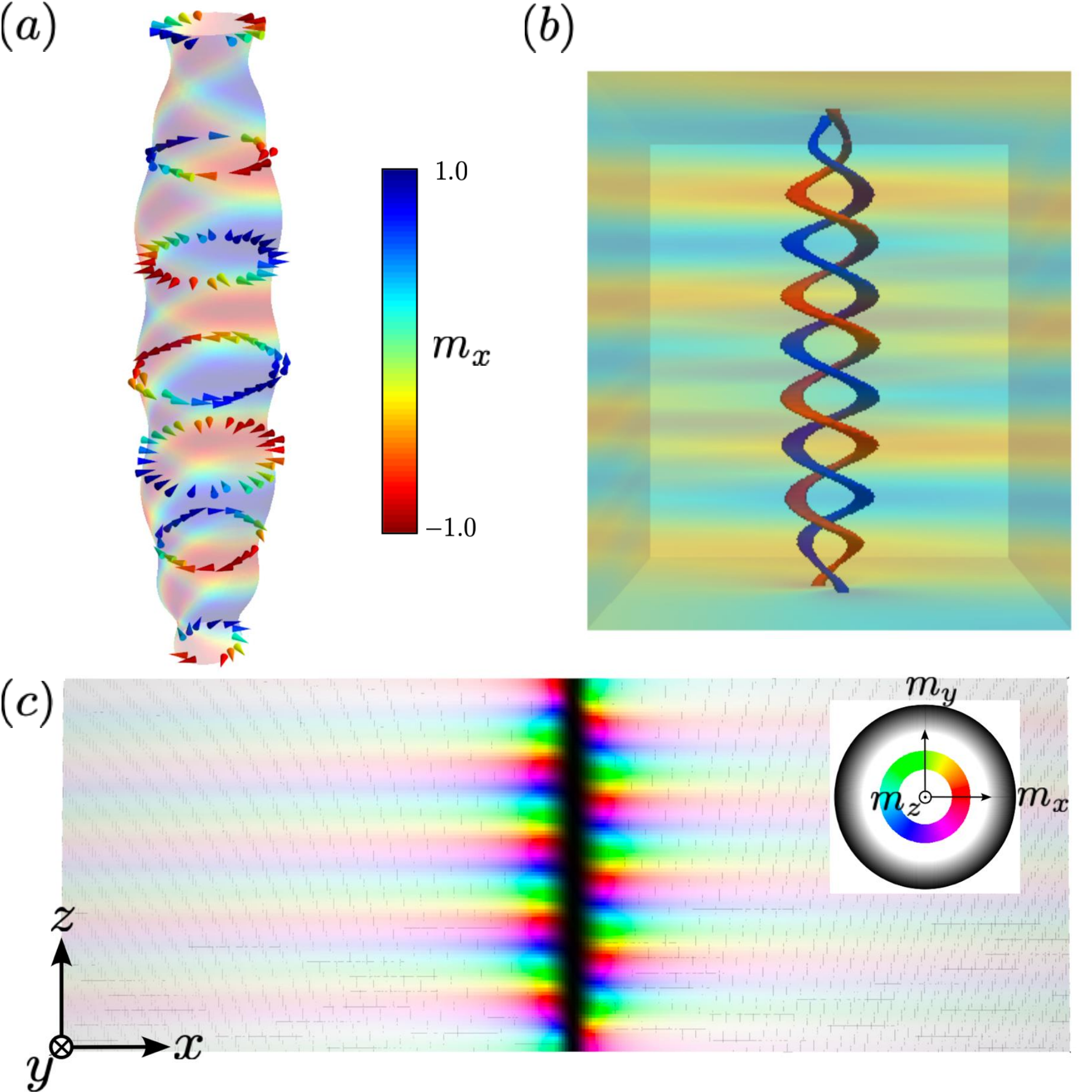}
\caption{Stabilized TSkT, obtained by numerical simulation, in a nano cylinder of radius $R=150$ nm, thickness $L_{z}=100$ nm, and external magnetic field $B_{z}=0.2$ T. (a) Magnetization field in the iso-surface $m_{z}=0$. (b) Pre-images of $\boldsymbol{m}=\hat{\boldsymbol{x}}$ (red) and $\boldsymbol{m}=-\hat{\boldsymbol{x}}$  (blue). (c) Horizontal midplane cross section in the $xz$-plane of the TSkT. 
}
\label{fig1}
\end{figure}

{Recently, Psaroudaki \textit{et al.} \cite{Psaroudaki2021} proposed the realization of skyrmion qubits in frustrated magnets based on the quantization of the helicity of the skyrmion. Thereby, the control of the helicity dynamics becomes a relevant feature to study such effects. In this context, a particularly intriguing extension of conventional skyrmion tubes is represented by twisted skyrmion tubes (TSkTs), which exhibit additional topological and geometrical properties. TSkTs represent a novel class of topological textures in magnetic materials consisting of a set of coplanar 2D skyrmions with varying helicity, as shown in Fig. \ref{fig1}. TSkTs with a single $\pi$ rotation of helicity have been reported to appear in chiral magnets \cite{Leonov2016,Yokota2021}, as a result of dipole-dipole interactions in centrosymmetric nanomagnets \cite{Kong-PRB}, or embedded within the conical state background in helimagnets \cite{Azhar2022,Leonov2021,Brearton2022}. However, the stability of TSkTs with twist angles multiples of $2\pi$ in a uniform state background remains elusive. Potential candidates for hosting TSkTs include the low-compound Pb$_{2}$VO(PO$_{4}$)$_{2}$  \cite{Kaul2004}, and GdRu$_2$Si$_2$ \cite{Khanh2020}.  

In this Letter, we investigate the stabilization and dynamics of TSkTs with generic twist angles in frustrated magnets with NNN exchange interactions. Models for frustrated magnets that feature competing next-nearest-neighbor (NNN) exchange interactions have been employed to stabilize localized topological textures such as skyrmions  \cite{Lin2016,Leonov2015,Hayami2021}, bimerons in monolayer systems \cite{Zhang2020}, and magnetic hopfions in 3D magnets \cite{Sutcliffe2017,Rybakov2022,Sallermann2023,Bogolubsky1988,Zhang2023}. We show that TSkTs possess a non-vanishing Hopf index $\mathcal{Q}_{H}$, similar to Hopfion rings \cite{Zheng2023}, combining the topological properties of skyrmions and hopfions. Interestingly, we find that the Hopf index is proportional to the system's thickness, making it generally non-integer and relatable to fractional Hopfions \cite{Sutcliffe2017,Rybakov2022,Sallermann2023, Bogolubsky1988,Zhang2023}. In addition, we show that the dynamics of TSkTs can by driven by spin-orbit torque interactions, similar to 2D skyrmions in frustrated magnets \cite{Knapman2024,Zhang2017,Yao2020,ZhangXia2023}. From this effect, we propose a dynamical mechanism to design nano-bateries.
 


{\it Theoretical Model.--} We consider a three-dimensional spin system defined on a cubic lattice, which is modeled by the energy density $\mathcal{E} =\mathcal{E}_{\text {ex }} + \mathcal{E}_{Z} + \mathcal{E}_{\mathrm{DDI}}$, where $\mathcal{E}_{Z}=  \vv{B}_{\mathrm{ext}}\cdot \boldsymbol{m}$ is the Zeeman energy by an external magnetic field $ \boldsymbol{B}_{\mathrm{ext}}$ applied along $z$-axis; the energy from the dipole-dipole interaction (DDI) is $\mathcal{E}_{\mathrm{DDI}}=(\mu_{0}\boldsymbol{H}_{\mathrm{dip}} \cdot \boldsymbol{m})/2$; and the contributions from 
Heisenberg exchange interaction is modeled up to four atomic shells of nearest-neighbor atoms, $\mathcal{E}_{\text {ex }}=-\sum_{s=1}^4 \sum_{\langle i, j\rangle_{s}} J_{s} \,\boldsymbol{m}_i \cdot \boldsymbol{m}_j$,
where the sum $\langle i, j\rangle_{s}$ runs over the sites $i$, $j$ which lie in each others $s$th-nearest-neighbor shell. For the exchange coupling strengths of the four shells of nearest-neighbor atoms, we set in this study the exchange values \cite{Kasuya1956,She2013}: $J_{1} = 93.6$ meV ferromagnetic interaction (FM) for the first shell, and $J_{2}=-0.1 J_{1}, J_{3}=-0.085 J_{1}, J_{4} = -0.068 J_{1}$ antiferromagnetic interaction (AFM) for the next shells. The presence of long range AFM exchange can be explained from the RKKY coupling model \cite{Kasuya1956,She2013}. In the continuum limit 
the exchange energy can be written as \cite{Rybakov2022}, $\mathcal{E}_{ex}= \sum_{\mu} A (\partial_{\mu} \boldsymbol{m})^{2}   +\sum_{\mu, \nu \neq \mu} C_{1} (\partial^{2}_{\mu \mu} \boldsymbol{m}-\partial^{2}_{\nu \nu} \boldsymbol{m})^2+C_{2} (\partial^{2}_{\mu \nu} \boldsymbol{m})^2$, being $\mu,\nu=\{x,y, z\}$, with the constants $A$, $C_1$ and $C_2$, and the competition among the FM and AFM exchange interactions determine the stability of non-collinear magnetic states.

Let us consider a magnetic field $\boldsymbol{B}_{\mathrm{ext}}=B_{z}\hat{\boldsymbol{z}}$, with $B_{z}>0$. 
We distinguish two scenarios: ferromagnet ($A>0$) and frustrated magnet ($A<0$, $C_{1}>0,C_{2}>0$), with different magnetic phases. For positive $A$, the ground state of the system is the uniform state $\boldsymbol{m}= \hat{\boldsymbol{z}}$. On the other hand, for negative $A$, and if it holds the condition $B_{z}<B_{c}=A^{2}/(8M_{s}C_{1})+\mu_{0}M_{s}$, where $B_{c}$ corresponds to the saturation field of the system (note that $\mu_{0}M_{s}$ takes into account the demagnetization field), then the ground state is determined by the conical state $\boldsymbol{m}_{C}(\boldsymbol{r})=(\cos(\boldsymbol{\kappa}\cdot \boldsymbol{r})\sin\theta,\sin(\boldsymbol{\kappa}\cdot \boldsymbol{r})\sin\theta,\cos\theta)$,  where the wave number $\boldsymbol{\kappa}=\kappa \hat{\boldsymbol{z}}$ and the angle $\theta$ are given by
\begin{align}
    \kappa=\frac{1}{2} \left (  \frac{|A|}{2 C_{1}}\right )^{1/2}, \quad \theta=\arccos \left ( \frac{B_{z}}{B_{c}}\right ). \label{eq:kz_Bc}
\end{align}
The relation between the exchange couplings $J_{1,2,3,4}$ and the stiffness constants are related by  $A=a^{-1}(J_{1}/2  + 2 J_{2} + 2  J_{3} + 2 J_{4})$, $C_{1}=-a (J_{1}/96 + J_{2}/24 + J_{3}/24 + J_{4}/6)$, and $C_{2}=-a (J_{1}/48 + J_{2}/3 + 7 J_{3}/12 + J_{4}/3)$ \cite{Rybakov2022,Sallermann2023}, where $a$ is the lattice constant. In this work, by substituting the exchange couplings $J_{s}$, we have $A= -6\cdot10^{-3} a^{-1}J_{1}$, $C_{1}=8.6\cdot 10^{-3} a J_{1}$ and $C_{2}=8.5\cdot10^{-2} aJ_{1}$, hence the system considered is a frustrated magnet ($A<0,C_{1}>0,C_{2}>0$). From Eq. (\ref{eq:kz_Bc}), we obtain $\kappa= 0.3$ nm$^{-1}$, hence the wavelength of the conical phase $\lambda_{c}=2\pi/\kappa =21.3$ nm, whereas the critical field provides $B_{c}= 0.2$ T. These results are in good agreement with numerical simulations performed using MuMax3 \cite{Mumax3} (discussed below), supporting the reliability of the analytical predictions.
\begin{figure}[!htb]
  \centering
\includegraphics[width=\linewidth]{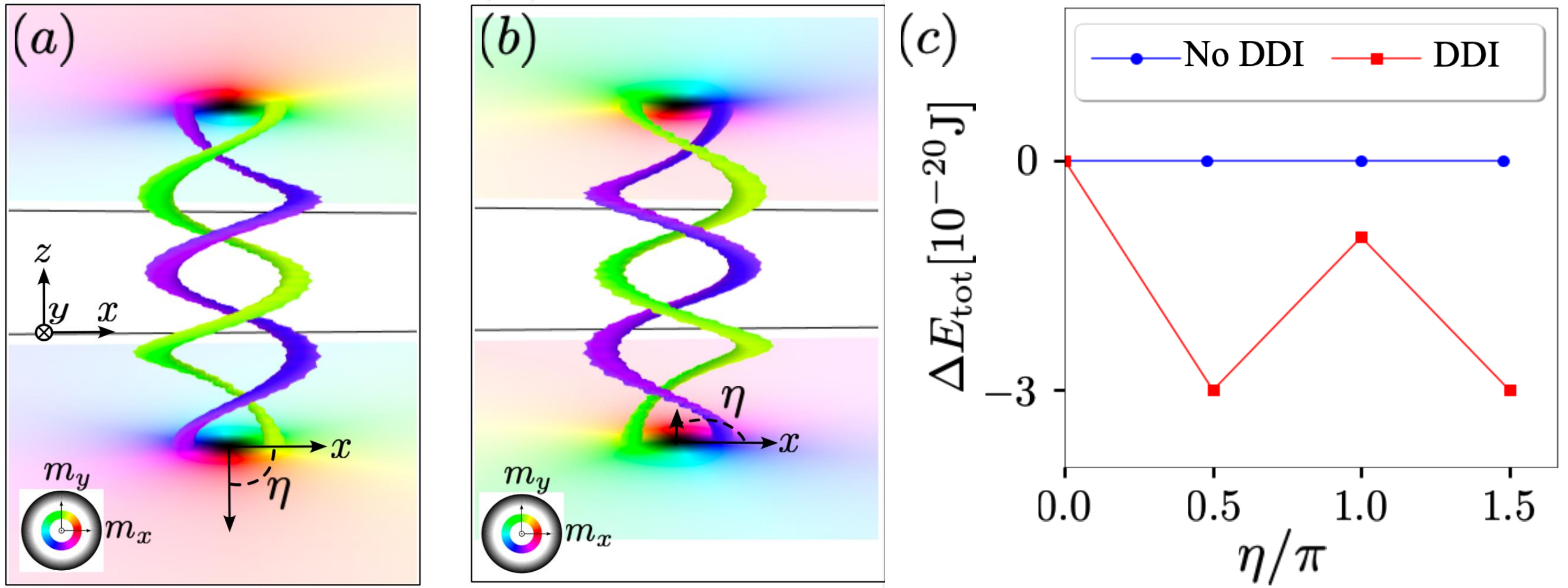}
\caption{TSkT with (a) anticlockwise chirality, where $\eta=\pi/2$ and Hopf index $\mathcal{Q}_{H}=-2$, and (b) clockwise chirality with $\eta=3\pi/2$ and Hopf index $\mathcal{Q}_{H}=2$. Preimages showing $\boldsymbol{m}=\hat{\boldsymbol{y}}$ (green) and $\boldsymbol{m}=-\hat{\boldsymbol{y}}$ (blue) components for a stabilized TSkT in a box of size $L_{x}=L_{y}=300$ nm and thickness film twice the period of the conical phase, $L_{z}=46 \ \mathrm{nm} \approx 2\lambda_{c}$. (c) Energy of a TSkT stabilized by micromagnetic simulations as a function of the helicity, with $\Delta E_{\mathrm{tot}}=E_{\mathrm{tot}}(\gamma)-E_{\mathrm{tot}}(0)$ for the cases: without DDI (blue line) and with DDI (red line).}
\label{fig:fig2}
\end{figure}

To show that TSkT are meta-stable states of frustrated magnets, we define $R$ as the characteristic size of the skyrmion tube. The analysis of the energy density allows us to state that the energy scales as $E(R)\approx A+ (C_{1}+C_{2})/{R^{2}} + B_{z} R^{2}$. Therefore, the energy reaches the minimum at $R=((C_{1}+C_{2})/B_{z})^{1/4}$, and the condition $C_{1}+C_{2}>0$ prevents the collapse of the skyrmion tube. To describe the TSkT, it is convenient to parametrize the magnetization field in cylindrical coordinates as $\boldsymbol{m}
=(\cos(\Phi)\sin(\Theta),\sin(\Phi)\sin(\Theta),\cos(\Theta))$. Thus, the spin configuration of a TSkT is well approximated by the ansatz: $\Theta= f(\rho)$, where $f$ is a monotonic function that satisfy the conditions $f(0)=\pi$, and $f(\infty)=0$. Additionaly, $\Phi= Q_{v}\phi+ \varphi(z)$, where $Q_{v}=\oint \partial_{i}\Phi \ d x^{i}/2\pi$ stands for the vorticity number. 
Since the terms $(\partial_{zx}^{2}\boldsymbol{m})^{2},(\partial_{zy}^{2}\boldsymbol{m})^{2}$ decay as $(\partial_{zx}^{2}\boldsymbol{m})^{2}\sim \rho^{-2}(\partial_{\mu}\boldsymbol{m})^{2}\ll (\partial_{\mu}\boldsymbol{m})^{2}$,  we can neglect the contribution of terms proportional to the coefficient $C_{2}$ in the energy functional. In this context, the effective energy of the TSkT is given by $E[\varphi]= E_{0}+ c_{0}\int \left[A(\varphi')^{2} + C_{1} (\varphi')^{4}+ C_{1} (\varphi'')^{2} \right] dz$, where $E_{0}$ encodes the energy of a skyrmion tube without twist ($\varphi=0$), and $c_{0}=\int_{0}^{\infty} \sin^{2}f(\rho) \rho d\rho$.  The Euler-Lagrange equation of $E[\varphi]$ leads to $A \varphi' + 2C_{1}(\varphi')^{3} - C_{1} \varphi''' =0$, which admits the solution $\varphi(z)=\pm\kappa z+\eta$,  where $\kappa= (|A|/(2C_{1}))^{1/2}/2$ corresponds to the wave number of the conical phase, given at Eq. (\ref{eq:kz_Bc}), and the parameter $\eta$ corresponds to the helicity of the TSkT. The sign $+$ ($-$) corresponds to an anticlockwise (clockwise) helicity, see Fig. \ref{fig:fig2}(a) and (b). Similar to 2D skyrmions, the helicity $\eta$ parametrizes the global symmetry group: $\boldsymbol{m}(\boldsymbol{r})\to \boldsymbol{m}'(\boldsymbol{r})=\mathcal{R}^{-1}_z[\eta]\boldsymbol{m}(\boldsymbol{r}) $, where $\mathcal{R}_z[\eta]$ is a rotation by an angle $\eta$ around the $z$-axis.
Note that the energy is invariant 
under such transformation in absence of DDI, and it becomes a Goldstone mode of the TSkT. On the other hand, the dipolar energy $\mathcal{E}_{\mathrm{DDI}}$ breaks the energetic degeneracy, yielding energy minima}. Fig. \ref{fig:fig2}(c) displays the energy difference $\Delta E_{\mathrm{tot}}= E_{\mathrm{tot}}(\gamma)-E_{\mathrm{tot}}(\gamma=0)$, where we observe that DDI effect favors the Bloch-type skyrmion helicities $\gamma=\pm \pi/2$, which can understood due to the skyrmion energy dependence on the helicity as $E_{\mathrm{DDI}} \propto -\sin(2\eta)$ \cite{Zhang2017}. 

Now, let us analyze the topological nature of TSkTs in frustrated magnets. Thus, we first determine the topological charge $\mathcal{Q}_{Sk}=\int_{z=z_{0}} \boldsymbol{m}\cdot \left(\partial_{x}\boldsymbol{m} \times \partial_{y}\boldsymbol{m}\right)d^{2}x/4\pi$, thought an arbitrary $xy$-cross-section along the $z=z_{0}$ plane of the considered system.
Since the topological charge is a topological invariant, $\mathcal{Q}_{Sk}$ does not depend on $ z_0$  in the absence of Bloch points within the system \cite{Koshibae2020}. In fact, we obtain $\mathcal{Q}_{Sk}=\int_{z=z_{0}} \boldsymbol{B}^{e}\cdot \hat{\boldsymbol{z}} \ d^{2}x/4\pi=-Q_{v}$ for all values of $z_{0}$. Here, $\boldsymbol{B}^{e}$ is an emergent magnetic field generated by the TSkT, given by  $(\boldsymbol{B}^{e})_{k}= \frac{1}{2}\epsilon_{ijk}\boldsymbol{m}\cdot \left(\partial_{i}\boldsymbol{m} \times \partial_{j}\boldsymbol{m}\right)$, from which we can define a vector potential $\boldsymbol{A}^{e}$. By using the TSkT solution, we have that
\begin{align}
\boldsymbol{B}^{e}&= -\kappa\sin(f)f'(\rho) \ \hat{\boldsymbol{\phi}} + \frac{Q_{v}}{\rho}\sin(f) f'(\rho) \ \hat{\boldsymbol{z}}, \label{eq:Bem}   \\ 
\boldsymbol{A}^{e}&= -\frac{Q_{v}}{\rho}(1+\cos(f) ) \ \hat{\boldsymbol{\phi}} + \kappa (1-\cos(f))\ \hat{\boldsymbol{z}},
\label{eq:Aem}
\end{align}
whose derivation is detailed at Supplemental Material (SM). The emergent field $\boldsymbol{B}^{e}$ is responsible for the topological Hall effect of electrons by skyrmions \cite{Neubauer2009, Hamamoto2015, Schulz2012}. The emergent magnetic field is displayed at SM, corresponding to the stabilized TSkT shown at Fig. \ref{fig1}. Although one observes modulation of the field as it approaches the top and bottom layers due to the surface magnetic charges, it is correctly described by Eq. (\ref{eq:Bem}) within the bulk of the system.

\begin{figure}[!htb]
  \centering
\includegraphics[width=\linewidth]{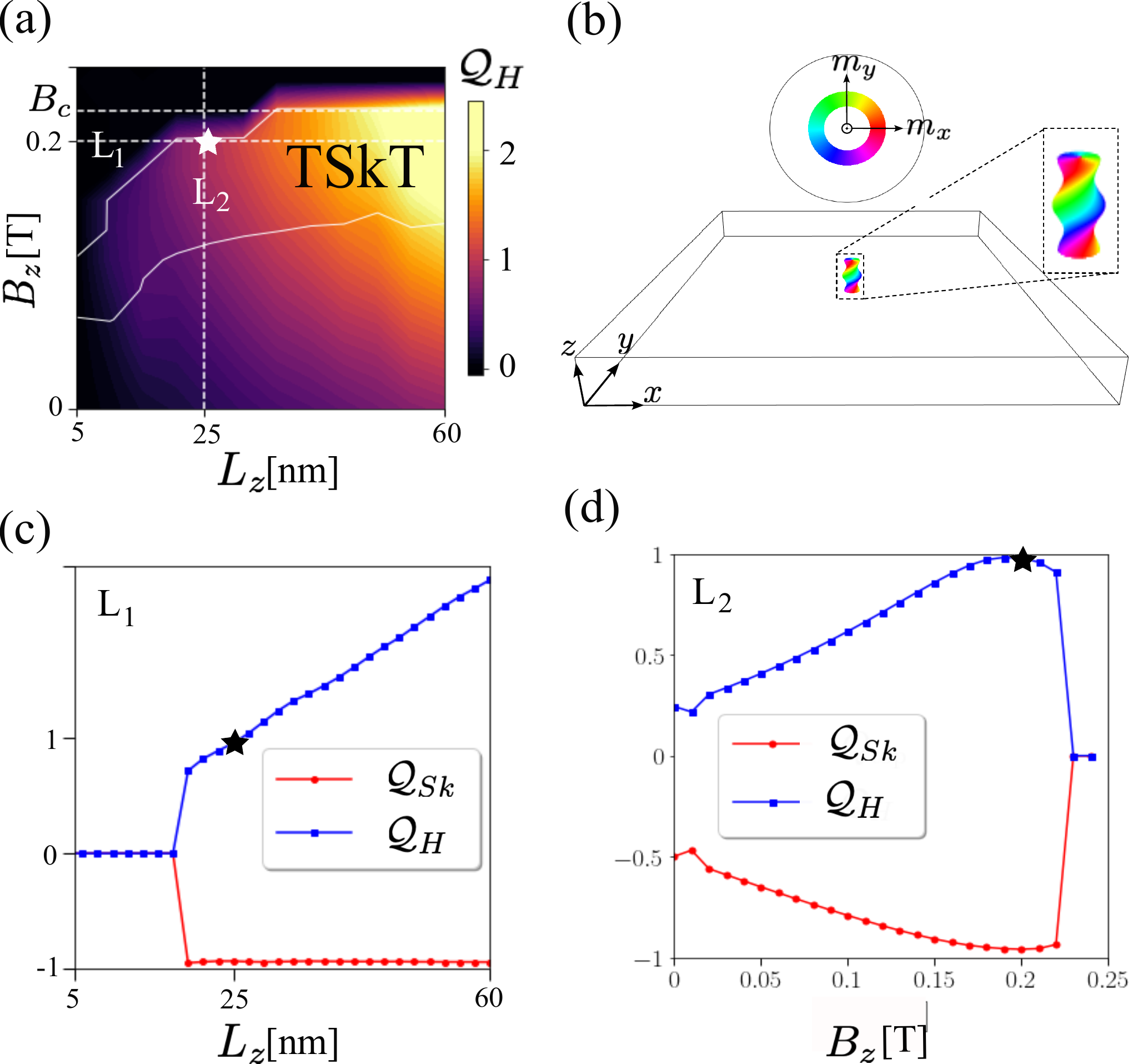}
\caption{(a) Stability diagram of the TSkT as a function of the external magnetic field $B_{z}$ and the thickness of the film $L_{z}$. Enclosed region corresponds to the value of the skyrmion charge between $-1\leq \mathcal{Q}_{Sk}<-0.9$. (b) Isosurface, $m_{z}=0$, representing the magnetization $(m_{x},m_{y})$ of a stabilized twisted skyrmion tube in a box of size $L_{x}=L_{y}=300$ nm, and $L_{z}=25\ \mathrm{nm} \approx \lambda_{c}$ ($Q_{H}\approx 1$) equal to the wave length of the conical state. (c),(d) Skyrmion topological charge $\mathcal{Q}_{Sk}$ (red line) and hopf index $\mathcal{Q}_{H}$ (blue line) as a function of $B_{z}$ and $L_{z}$, respectively.}
\label{fig:QvsBz}
\end{figure}

Now, analyzing the 3D nature of the TSkT, we determine its Hopf index \cite{Hopf1931}, defined by the integral
\begin{equation}\label{eq:QH}
\mathcal{Q}_{H} = -\frac{1}{(4 \pi)^2} \int_V \boldsymbol{B}^{e} \cdot \boldsymbol{A}^{e} \ d^{3}x,
\end{equation}
which determines how knotted pairs of pre-images of $\boldsymbol{m}$ are. Using Eqs. (\ref{eq:Bem}) and (\ref{eq:Aem}) we find the Hopf index obeys $\mathcal{Q}_{H} = { Q_{v}\,\kappa  L_{z}}/{2\pi}$, hence $\mathcal{Q}_{H}=-\mathcal{Q}_{Sk}N_{\mathrm{twist}}$, where $N_{\mathrm{twist}}=\kappa L_{z}/2\pi$ stands for the number of twists of the TSkT (see Fig. \ref{fig:QvsBz}). For our numerical computation, we evaluate the vector potential and the Hopf index by considering the gauge fixing \cite{Knapman2025}, $\boldsymbol{A}^{e}(x,y,z)= \int_{0}^{y}(\boldsymbol{B}^{e}(x,y',z) \times  \hat{\boldsymbol{y}})\ dy'$,
and using Eq. (\ref{eq:QH}), respectively. In particular, for the stabilized TSkT of Fig. \ref{fig1}, it leads to $\mathcal{Q}_{H}\approx 4.3$, whereas the theoretical prediction  $\mathcal{Q}_{H}\approx 4.5$, as we can see from Fig. \ref{fig:QvsBz} (d). The Hopf invariant is defined for smooth, continuous, and unit-norm vector fields that is classified by homotopy group $\pi_3(S^2) = \mathbb{Z}$. Thus, the emergence of a semi-integer values, suggests that the texture is not globally well-defined as a topological soliton. In our case, non-integers $\mathcal{Q}_{H}$ arises due to a discontinuity at the system boundary. Due to the lack of topological protection, the configuration may eventually decay into a stable state with $Q_H = 4$, provided the system can overcome the associated energy barriers. Interestingly, the emergent magnetic field $\boldsymbol{B}^{e}$ induces a non-vanishing toroidal moment  $\boldsymbol{\mathcal{T}}= \int_{V} (\boldsymbol{r}\times \boldsymbol{B}^{e} ) d^{3}r$, that satisfy $\boldsymbol{\mathcal{T}}= \mathcal{T}_{z}\,\hat{\boldsymbol{z}}$, where $\mathcal{T}_{z}= 2\pi \kappa L_{z}\int_{0}^{\infty} (\cos[f(\rho)]-1) \rho \ d\rho \approx   4\pi \kappa  L_{z} R^{2}$,
with $R$ representing the radius of the TSkT. Nonzero toroidal moments reflects a chiral distribution of $\boldsymbol{B}^{e}$ that breaks spatial inversion and time-reversal symmetries, leading to nonreciprocal responses, such as asymmetric spin-wave propagation or directional-dependent transport \cite{Saji2023}. In addition, the dependence of the $\mathcal{T}_z$ on the system thickness provides a tunable parameter for spintronic applications.

{\it Micromagnetic Simulations.--} The simulations were performed using the GPU-accelerated micromagnetic software package  $\mathrm{Mumax}^{3}$ \cite{Mumax3}, which solve Landau-Lifschitz-Gilbert (LLG) equation,
\begin{align}
\partial_t \bm{m}=-\gamma \mathbf{m} \times \bm{H}_{\mathrm{eff}}+\alpha \bm{m} \times \partial_t \bm{m},
\end{align}
where $\gamma$ is the electron gyromagnetic ratio, $\alpha$ is the Gilbert damping constant, and the effective field is given by $\mu_{0} {\bm H}_{\text {eff }}= \sum_{s}\sum_{\langle i, j\rangle_{s}}{J_{s}} \boldsymbol{m}_{j}/M_{s} + \boldsymbol{B}_{ext}+ \mu_0 {\bm H}_{\text {dip}},
$
where the first term on the right-hand side takes into account the contribution of the exchange Heisenberg of the $s$th-nearest-neighbor shell. The geometry dimensions are $L_{x}=300$ nm, $L_{y}= 300$ nm and $L_{z}$ in the range $5$ nm to $100$ nm. The cell size is $a=1$ nm. The magnetic parameters used in the simulations are as follows: exchange stiffness constant $A_{\mathrm{ex}}= 15$ [pJ/m], saturation magnetization $M_{\mathrm{sat}}= 100$ [kA/m], and the damping constant is set to be $\alpha=0.1$. The initial configuration corresponds to a phase modulated skyrmion tube with the domain-wall-like ansatz $\Theta(\rho,\phi,z)=\cos^{-1}\left ( \tanh(\frac{\rho-\rho_{0}}{w}) \right )$ and $\Phi(\rho,\phi,z) =Q_{v}\phi+ \kappa z+\eta$, where $Q_{v}=1$, $\rho_{0}=30$ nm, $\eta=\pi/2$, and $w=25$ nm, the wave number of the conical phase for the set of parameters above. To obtain the TSkT, the system is relaxed to its steady state. Our simulations yield $\lambda_{c} = 21$ nm, $B_{c} = 0.22$ T in the presence of DDI, and $B_{c} = 0.1$ T in its absence. These values are in excellent agreement with the analytical predictions, as previously anticipated. To calculate the radii $R$ of the TSkT, we compute the average distance of the points on the $m_{z}=0$ contour to the center along the $z$-axis. To compute the helicity, we determine the average angle $\theta=\cos^{-1}(\boldsymbol{m}\cdot \hat{\boldsymbol{\rho}})$ of the magnetization on the $\boldsymbol{m}$  with $\hat{\boldsymbol{\rho}}$ is the radial direction.

Figure \ref{fig:QvsBz} displays the average topological charge $\mathcal{Q}_{Sk}$ and the Hopfion index $\mathcal{Q}_{H}$ as a function of the external magnetic field and the film thickness $L_{z}$. We observe that as the external field $B_{z}$ increases, the in-plane texture smoothly transforms from a meron (half-skyrmion) with $\mathcal{Q}_{Sk}=-1/2$ to a skyrmion with $\mathcal{Q}_{Sk}=-1$ as it approaches the critical $B_{c}=0.2$ T.  TSkT remain stable for values of the external magnetic field in some range close to the critical field $0.17\ \mathrm{T}<B_{z}<0.22\ \mathrm{T}$, where the value of $\mathcal{Q}_{Sk}$ and $\mathcal{Q}_{H}$ approach $-1$ and $1$, respectively, as shown in Fig. \ref{fig:QvsBz} (b).  The magnetic phase diagram of the magnetic states corresponding to the stability of the TSkT for the frustrated magnet considered in this letter is displayed at Fig. \ref{fig:QvsBz} (a), for the film thickness $L_{z}$ and the applied magnetic field $B_{z}$.

{\it Current-driven dynamics.--} We now investigate the evolution dynamics of a TSkT driven by spin-orbit torque (SOT) \cite{Liu2012,Liu22012,Sinova2015,Amin2020}. Such an effect can be described effectively in the LLG equation by the torque, $\boldsymbol{\tau}_{\mathrm{SOT}}= -\tau_{\mathrm{FL}}\boldsymbol{m}\times \boldsymbol{\sigma} - \tau_{\mathrm{DL}}\boldsymbol{m}\times(\boldsymbol{m} \times \boldsymbol{\sigma})$, where $\boldsymbol{\sigma}$ encodes the spin polarization of the electric current. The factors ${\tau}_{\mathrm{DL}}$ and ${\tau}_{\mathrm{FL}}$ are the strengths of damping-like (DL) and field-like (FL) torques, respectively, which are material dependent coefficients related by $\tau_{\mathrm{FL}}=\chi \tau_{\mathrm{DL}}$ \cite{Sinova2015,Amin2020}, with $\chi$ a constant. Here, we consider the out-of-plane spin-polarization $\boldsymbol{\sigma}=\sigma_{z} \hat{\boldsymbol{z}}$ along the $z$-axis. This kind of spin-polarization has been achieved in the non-collinear antiferromagnets \cite{Nan2020,Hu2022}.

\begin{figure}[!htb]
  \centering
\includegraphics[width=\linewidth]{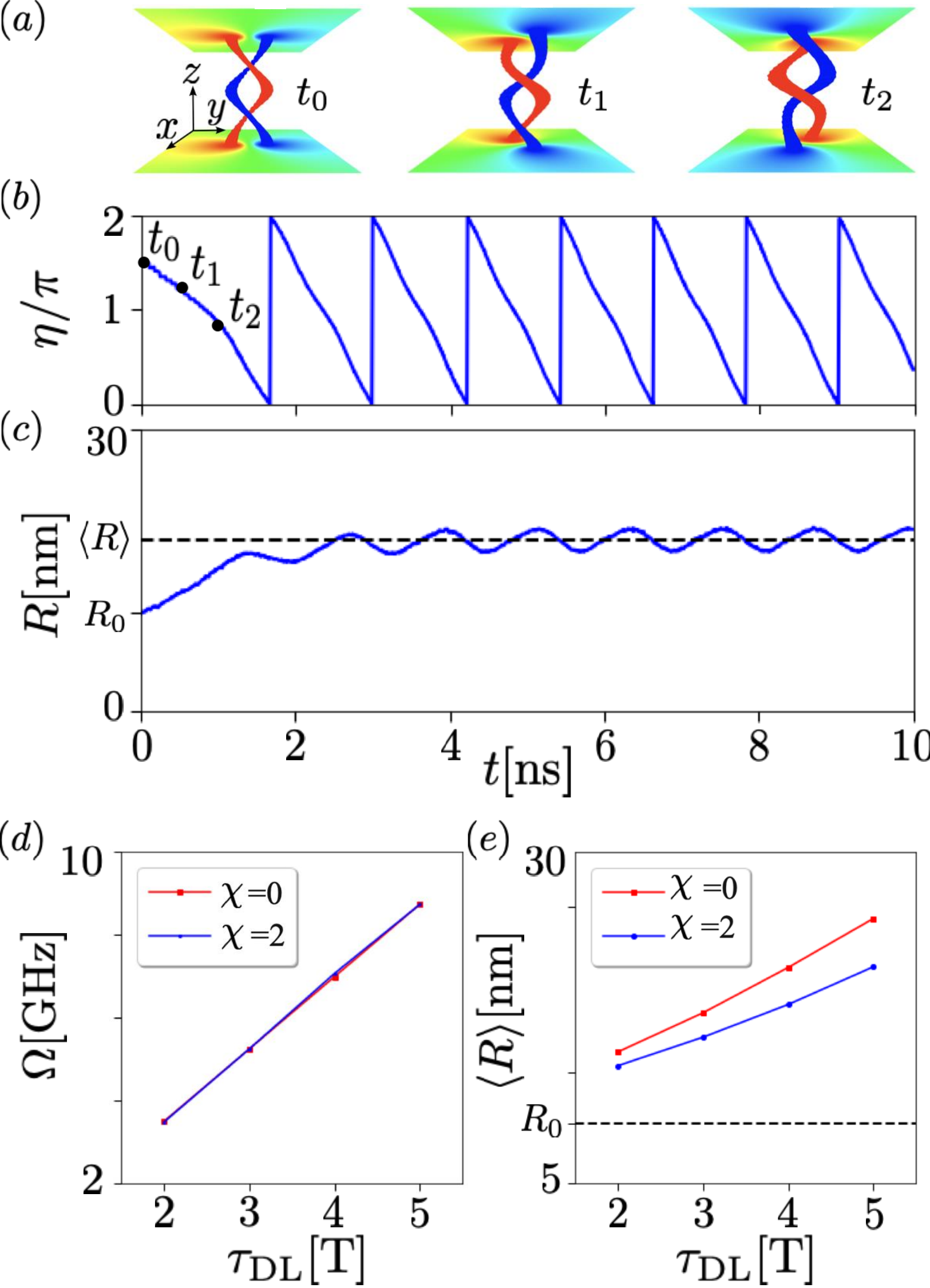}
\caption{Representation of the SOT-driven TSkT (with $\mathcal{Q}_{H}=1$). (a) Helicity dynamics of a TSkT driven by spin orbit torque with spin polarization $\boldsymbol{\sigma}=-\hat{\boldsymbol{z}}$, indicating a clockwise rotation.  (b) Time evolution of $R(t)$, and the steady state oscillations. (c) rotation of the pre-images $\boldsymbol{m}=\hat{\boldsymbol{x}}$ (red curve) and $\boldsymbol{m}=-\hat{\boldsymbol{x}}$ (blue curve). (d) Frequency of a TSkT shown at (b) as a function of $\tau_{\mathrm{DL}}$ with $\tau_{\mathrm{FL}}=\chi \tau_{\mathrm{DL}}$ for  $\chi=0$ (red line), $\chi=2$ (blue line). (e) Average radius of a TSkT as a function of the $\tau_{\mathrm{DL}}$ with $\tau_{\mathrm{FL}}=\chi \tau_{\mathrm{DL}}$ for  $\chi=0$ (red line), $\chi=2$ (blue line).}
\label{figure4}
\end{figure}

To describe effective dynamics, the TSkT is assumed to be a fixed shape object, allowing parameterize its motion in terms of the collective coordinates: $\xi(t)= (\varphi(t),R(t),X(t),Y(t))$, namely, the helicity ($\varphi$), the radius ($R$), and the position of the center of mass  $(X,Y)$. We remark that, as conventional skyrmions, the pairs $(\eta,R)$ and $(X,Y)$ are both canonically conjugate variables. Under this approach, the Thiele's equation models the effective quasiparticle motion equation of the TSkT:
\begin{equation}\label{eq:Thiele}
    -\mathcal{G}_{ij}\dot{\xi}^{j} + \alpha \ \mathcal{D}_{ij}\dot{\xi}^{j} = -\partial_{i}E + f^{\mathrm{SOT}}_{i},
\end{equation}
where $\mathcal{G}_{ij}= \int_{V}  \boldsymbol{m}\cdot \left(\partial_{\xi_{i}}\boldsymbol{m} \times \partial_{\xi_{j}}\boldsymbol{m}\right)\ d^{3}r $ and $\mathcal{D}_{ij}=\int_{V} \partial_{\xi_{i}}\boldsymbol{m}\cdot \partial_{\xi_{j}}\boldsymbol{m} \ d^{3}r$  stand for the gyrotropic and dissipative tensor, respectively; $E(R)$ is the TSkT energy as a function of $R$ (note that it does not depend on $\eta$ as we neglect the DDI energy contribution), and the last term $f^{\mathrm{SOT}}_{i}=-\int_{V} \partial_{\xi_{i}}\boldsymbol{m}\cdot \boldsymbol{\tau}_{\mathrm{SOT}} \ d^{3}r$ is the generalized force due to the SOT interaction. Due to the axial symmetry, the force $f^{\mathrm{SOT}}_{i}(R)$ depends only on $R$ (see SM for a more detailed derivation), and the matrix elements $\mathcal{G}_{i\mu}=0$ vanish for $i=\eta,R$ and $\mu=X,Y$. Moreover, one can show that the forces $f^{\mathrm{SOT}}_{X}=f^{\mathrm{SOT}}_{Y}=0$, and consequently, from Eq. (\ref{eq:Thiele}), we obtain $\dot{X}=\dot{Y}=0$. Thus, we can rewrite the Eq. (\ref{eq:Thiele}) in the form:
\begin{equation}\label{eq:eta_R_dynamics}
\begin{pmatrix}
\alpha \mathcal{D}_{\eta \eta}  & -\mathcal{G}_{R\eta}  \\
\mathcal{G}_{R\eta} &  \alpha \mathcal{D}_{R R}\end{pmatrix} 
\begin{pmatrix}
\dot{\eta} \\  \dot{R}
\end{pmatrix}= \begin{pmatrix}
f^{\mathrm{SOT}}_{\eta} \\ f^{\mathrm{SOT}}_{R} - E'(R) \end{pmatrix}.
\end{equation}
Now, let us consider the linear approximation of the TSkT dynamics around the equilibrium configuration, $R=R_{0}+\delta R$. Since $R_{0}$ is the equilibrium point, we have that $E'(R_{0})=0$, so remain terms up to first order in $\delta R$, we deduce that $\eta(t)$ rotates with constant angular frequency given by
\begin{align*}
\Omega = \frac{\mathcal{G}_{R\eta} f^{\mathrm{SOT}}_{R}+ \alpha \mathcal{D}_{RR} f^{\mathrm{SOT}}_{\eta}}{\mathcal{G}_{R\eta}^{2}+\alpha^{2}\mathcal{D}_{RR}\mathcal{D}_{\eta \eta}} \approx  -\sigma_{z}\left(\tau_{\mathrm{DL}}-\alpha\tau_{\mathrm{FL}}\right),
\end{align*}
where we have used that matrix elements $\mathcal{G}_{R\eta}$, $\mathcal{D}_{RR}$, $\mathcal{D}_{\eta \eta}$, $f^{\mathrm{SOT}}_{\eta}$, and $f^{\mathrm{SOT}}_{R}$ are linear functions in $R$, whose coefficients are computed in the SM. Figure \ref{figure4} illustrates the time evolution of $(\eta,R)$ obtained by micromagnetic simulations in Mumax3 for a TSkT with $\mathcal{Q}_{H}=1$, illustrated in the Fig. \ref{fig:QvsBz} (b). According to the numerical simulations, after some transient time, the asymptotic dynamics reaches a time-periodic steady state where the helicity $\eta(t)$ describes a clockwise rotation motion, whereas $R(t)$ oscillates around an time-average radius $\left\langle R \right\rangle$ as $t\to \infty$. The dependency of frequency and average radius of a TSkT as a function of $\tau_{\text{DL}}$ is shown at Fig. \ref{figure4}(d) and \ref{figure4}(e), respectively.

Finally, according to Faraday’s law, a time dependent spin texture generates an emergent electric field $\boldsymbol{E}_{\mathrm{em}}$ given by $(\boldsymbol{E}_{\mathrm{em}})_{i}= {\hbar} \boldsymbol{m}\cdot \left(\partial_{i}\boldsymbol{m} \times \partial_{t}\boldsymbol{m}\right)/q_{e}$ \cite{Barnes2007}. Using the ansatz for the TSkT dynamics: $\Theta= f(\rho-R(t)), \Phi= Q_{v}\phi+\kappa z+\Omega t+\eta_{0}$, we obtain $\boldsymbol{E}_{\mathrm{em}}= {\hbar}  \partial_{\rho}\Theta \sin (\Theta)( -\Omega \hat{\boldsymbol{\rho}} + \dot{R} Q_{v}\hat{\boldsymbol{\phi}} +\dot{R}\kappa \hat{\boldsymbol{z}})/{q_{e}}$. The resulting spin-motive force $q_{e}\boldsymbol{E}_{\mathrm{em}}$ acting on the electrons is radially oriented and with a signature the same as $\Omega$ (and $\sigma_{z}$). The spin-motive force acting on the conduction electrons generates an average DC electric voltage given by $\bar{V}_{\mathrm{DC}}= \int_{0}^{\infty} \boldsymbol{E}_{\mathrm{em}} \cdot d\boldsymbol{\rho}= {2\hbar}\Omega/{q_{e}}$,
and therefore, a SOT-driven TSkT behaves as a battery with capacitance $C=q_{e}^{2}/(2\hbar \Omega)$, which might offers applications for spintronics.  

{\it Conclusions.--} We have established the stability of TSkT in frustrated magnets and discussed its dynamics driven by an electric current. Our results evidence that TSkTs are important candidates to be used in applications of technology for low-power storage nano-devices and nano-batteries with enhanced control. Additionally, these topological textures promise to enable building blocks of logic-gate devices and neuromorphic computing systems. By presenting these new solutions, we aim to expand the theoretical framework and experimental toolbox for studying magnetic TSkT, setting the ground for future innovations in spintronics. Experimental detection and tracking of TSkT could be performed using Lorentz transmission electron microscopy (TEM) and electron holography. The helicity can be calculated from TEM measurements or using scanning transmission X-ray microscopy (STXM), magneto-optical Kerr effect (MOKE) microscopy, and magnetic force microscopy (MFM). Specifically, the electronic scattering problem by a TSkT can be addressed similarly to the hopfion case \cite{Pershoguba2021} since a non-vanishing toroidal magnetic moment $\boldsymbol{\mathcal{T}}$, leading to a viable method for the experimental detection of TSkTs using TEM.

{\it Acknowledgments.-} C.S. thanks the financial support provided by the ANID National Doctoral Scholarship Nº21210450. A.S.N. acknowledges funding from Fondecyt Regular 1230515. R.E.T. thanks funding from Fondecyt Regular 1230747. E.S. acknowledges support from Dicyt-USACH 042331SD. V.L.C.-S. acknowledges CNPq, Fapemig, INCT/CNPq - Spintr{{\^o}nica e Nanoestruturas Magn\'eticas Avan\c{c}adas
(INCT-SpinNanoMag), and Rede Mineira de Nanomagnetismo/FAPEMIG. 

\bibliography{Biblio}

\clearpage
\onecolumngrid
\section{Supplemental Material}
In this Supplemental Material, we show explicitly details of calculations of parameters involved in the dynamics of the TSkT and its emergent magnetic field.

\subsection{Analytical evaluation of $\mathcal{G}_{ij},\mathcal{D}_{ij},f^{\mathrm{SOT}}_{i}$ }

In this section we focus on the analytically calculate the  quantities $\mathcal{G}_{ij},\mathcal{D}_{ij},f^{\mathrm{SOT}}_{i}$ using this ansatz of TSkT as a function of the collective coordinates $\xi=(\eta,R,X,Y)$ in the form:  $\Theta(x,y,z)= f(\rho-R)$ where $f$ is function such that $f(0)=-\pi, \ f(\infty)=0$, and $\Phi(x,y,z)=Q_{v}\phi+ \kappa z+\eta$ , where $\phi=\tan^{-1}((y-Y)/(x-X)),\rho=((x-X)^{2}+(y-Y)^{2})^{1/2}$. Furthermore, we approximate the skyrmion profile by using the domain-wall like ansatz: 
\begin{equation*}
\Theta=\cos^{-1}\left ( \tanh(\frac{\rho-R}{w}) \right )
\end{equation*}
with $w\ll R$. In particular we have that $\partial_{R}\Theta=-\partial_{\rho}\Theta$. Using the chain rule, we obtain:

\begin{equation}\label{eq:partial_m}
    \partial_{\xi}\boldsymbol{m}= \partial_{\xi}\Theta \ \boldsymbol{e}_{\Theta} + \sin(\Theta)\partial_{\xi}\Phi\  \boldsymbol{e}_{\Phi}
\end{equation}
hence $\partial_{\eta}\boldsymbol{m}= \partial_{\Phi}\boldsymbol{m}= \sin(\Theta)\boldsymbol{e}_{\Phi}$ and $\partial_{R}\boldsymbol{m}= \partial_{R}\Theta \partial_{\Theta}\boldsymbol{m}=-\partial_{\rho}\Theta \boldsymbol{e}_{\Theta}$, where $(\boldsymbol{e}_{\Theta}, \boldsymbol{e}_{\Phi})$ are given by $\boldsymbol{e}_{\Theta}= (\cos(\Theta)\cos(\Phi),\cos(\Theta)\sin(\Phi),-\sin(\Theta))$ and $\boldsymbol{e}_{\Phi}=(-\sin(\Phi),\cos(\Phi),0)$. Using the above formulas, the gyrotropic tensor yields
\begin{align*}
\mathcal{G}_{R\eta} &= \int_{V}  \boldsymbol{m}\cdot \left(\partial_{R}\boldsymbol{m} \times \partial_{\eta}\boldsymbol{m}\right)\ d^{3}r = \int_{V}  \partial_{R}\Theta \sin(\Theta)\ \boldsymbol{m}\cdot \left(\boldsymbol{e}_{\Theta} \times \boldsymbol{e}_{\Phi}\right)\ d^{3}r = 2\pi L_{z}\int_{0}^{\infty}  \partial_{\rho}\Theta \sin(\Theta)\ \rho d\rho \approx -4 \pi L_{z}R
\end{align*}
whereas the dissipative tensor is given by,
\begin{align*}
\mathcal{D}_{RR} &= \int_{V}  (\partial_{R}\boldsymbol{m})^{2} \ d^{3}r =2\pi L_{z}\int_{0}^{\infty}  (\partial_{\rho}\Theta)^{2} \rho d\rho \approx 2 \pi L_{z}\frac{R}{w}\\
\mathcal{D}_{\eta \eta} &= \int_{V}(\partial_{\eta}\boldsymbol{m})^{2} \ d^{3}r =2\pi L_{z}\int_{0}^{\infty}  \sin^{2}(\Theta) \rho d\rho \approx 2 \pi L_{z}w R \\
\mathcal{D}_{R\eta} &= \int_{V}  \partial_{R}\boldsymbol{m}\cdot  \partial_{\eta}\boldsymbol{m} \ d^{3}r =0.
\end{align*}

Now, let us calculate the $X,Y$ elements matrix of the Gyrotropic. Using \ref{eq:partial_m}, we obtain:
\begin{align*}
\mathcal{G}_{\eta X} &= \int_{V}  \boldsymbol{m}\cdot \left(\partial_{\eta}\boldsymbol{m} \times \partial_{X}\boldsymbol{m}\right)\ d^{3}r = - L_{z}\int  \partial_{x}\Theta \sin(\Theta)\ d^{2}x =0 \\
\mathcal{D}_{\eta X} &= \int_{V}  \boldsymbol{m}\cdot \left(\partial_{\eta}\boldsymbol{m} \times \partial_{X}\boldsymbol{m}\right)\ d^{3}r =  - L_{z}\int  \partial_{x}\Theta \sin^{2}(\Theta)\ d^{2}x =0.
\end{align*}

Similarly, $ \mathcal{G}_{Y\eta}=0 , \mathcal{G}_{XR}=0, \mathcal{G}_{YR}=0, \mathcal{D}_{Y\eta}=0 , \mathcal{D}_{XR}=0, \mathcal{D}_{YR}=0$.  Therefore, the gyrotropic and dissipative tensors read:
\begin{align*}
\mathcal{G} &= \begin{pmatrix}
0 &  \mathcal{G}_{\eta R} & 0 & 0 \\
-\mathcal{G}_{\eta R} & 0 &  0 & 0 \\
0 & 0 &  0 &  \mathcal{G}_{XY} \\
0 & 0  & -\mathcal{G}_{XY}  & 0 \\
\end{pmatrix}, \\
\mathcal{D} &= \begin{pmatrix}
\mathcal{D}_{\eta \eta} &  0  & 0 & 0 \\
0 & \mathcal{D}_{R R} &  0 & 0 \\
0 & 0 &  \mathcal{D}_{XX} &  0 \\
0 & 0  & 0  & \mathcal{D}_{XX} \\
\end{pmatrix}.
\end{align*}

Let us consider the spin polarization $\boldsymbol{\sigma}=\sigma_{z}\hat{\boldsymbol{z}}$, then we separate the generalized force in the damping-like and field-like contributions to the SOT as $f^{\mathrm{SOT}}_{i}= f^{\mathrm{DL}}_{i}+f^{\mathrm{FL}}_{i}$. Using the previous ansatz, we obtain

\begin{align*}
f^{\mathrm{FL}}_{\eta} &= \sigma_{z} \tau_{\mathrm{FL}}\int_{V} \hat{\boldsymbol{z}}\cdot(\boldsymbol{m} \times \partial_{\eta}\boldsymbol{m}) \ d^{3}r = 2\pi \sigma_{z} \tau_{\mathrm{FL}}L_{z} \int_{0}^{\infty} \sin^{2}(\Theta)\rho d\rho 
\end{align*}
and $f^{\mathrm{FL}}_{R} = \sigma_{z} \tau_{\mathrm{DL}}\int_{V} \hat{\boldsymbol{z}}\cdot(\boldsymbol{m} \times \partial_{R}\boldsymbol{m}) \ d^{3}r= 0$. Similarly, we have that
\begin{align*}
f^{\mathrm{DL}}_{R} = 2\pi \sigma_{z} \tau_{\mathrm{DL}}L_{z}\int_{0}^{\infty} \partial_{R}\Theta \sin(\Theta)  \rho d\rho 
\end{align*}
and $f^{\mathrm{DL}}_{\eta}= 0$. Therefore, we can estimate \begin{align*}
f^{\mathrm{SOT}}_{\eta} & \approx 2\pi \sigma_{z} \tau_{\mathrm{FL}}L_{z} w R\\
f^{\mathrm{SOT}}_{R} & \approx -4\pi\sigma_{z} \tau_{\mathrm{DL}}L_{z}R.
\end{align*}
On the other hand, 
\begin{align*}
f^{\mathrm{FL}}_{X} &= -\sigma_{z} \tau_{\mathrm{FL}}\int_{V} \hat{\boldsymbol{z}}\cdot(\boldsymbol{m} \times \partial_{x}\boldsymbol{m}) \ d^{3}r = \sigma_{z} \tau_{\mathrm{FL}}L_{z} \int_{0}^{\infty} \sin(\Theta)\partial_{x}\Phi\ d^{2}x =0
\end{align*}
where we have used the identity $\partial_{X}\boldsymbol{m}=-\partial_{x}\boldsymbol{m}$. Similarly $f^{\mathrm{FL}}_{X}=f^{\mathrm{FL}}_{Y}=f^{\mathrm{DL}}_{Y}=0$.

\subsection{Emergent magnetic field of a TSkT}
We calculate the emergent magnetic field of a TSkT described by the ansatz given by $\boldsymbol{m}= (\cos(\Phi)\sin(\Theta),\sin(\Phi)\sin(\Theta),\cos(\Theta)$ where $\Theta(\rho,\phi,z)= f(\rho-R)$ where $f$ is function such that $f(0)=-\pi, \ f(\infty)=0$, and $\Phi(\rho,\phi,z)=Q_{v}\phi+ \kappa z+\eta$. Using the formula
$$
\partial_{i}\boldsymbol{m}= \partial_{i}\Theta \ \boldsymbol{e}_{\Theta} + \sin(\Theta)\partial_{i}\Phi\  \boldsymbol{e}_{\Phi}
$$
and taking the cross product, we obtain 
$$
\partial_{i}\boldsymbol{m} \times \partial_{j}\boldsymbol{m}= f'(\rho)\sin(f) \left ( \partial_{i}\Phi \partial_{j}\rho -  \partial_{j}\Phi \partial_{i}\rho \right ) \ \boldsymbol{m}\,.
$$
Therefore, we have that 
$$
(\boldsymbol{B}^{e})_{k}= \frac{1}{2}\epsilon_{ijk}\boldsymbol{m}\cdot (\partial_{i}\boldsymbol{m} \times \partial_{j}\boldsymbol{m})= f'(\rho)\sin(f)\epsilon_{ijk}\partial_{i}\Phi \partial_{j}\rho 
$$
now using that $\partial_{i}\Phi=Q_{v}\partial_{i}\phi+ \kappa \partial_{i}z$, and the canonical vector frame $\hat{\boldsymbol{\rho}},\hat{\boldsymbol{\phi}},\hat{\boldsymbol{z}}$, we obtain the formula \ref{eq:Bem}.  A straightforward derivation we have that $\nabla \times \boldsymbol{A}^{e}= \boldsymbol{B}^{e}$. Using the formulas (3) and (4) in the main text, the Hopfion index yields 
\begin{align*}
\mathcal{Q}_{H} =\frac{Q_{v}\,\kappa}{4 \pi} \int_{0}^{L_{z}} \int_{0}^{\infty} {\frac{\sin [f(\rho)] }{\rho}f'(\rho) \rho d \rho  d z }= \frac{ Q_{v}\,\kappa  L_{z}}{2\pi}.
\end{align*}
\begin{figure}[!htb]
  \centering
\includegraphics[width=0.3\textwidth]{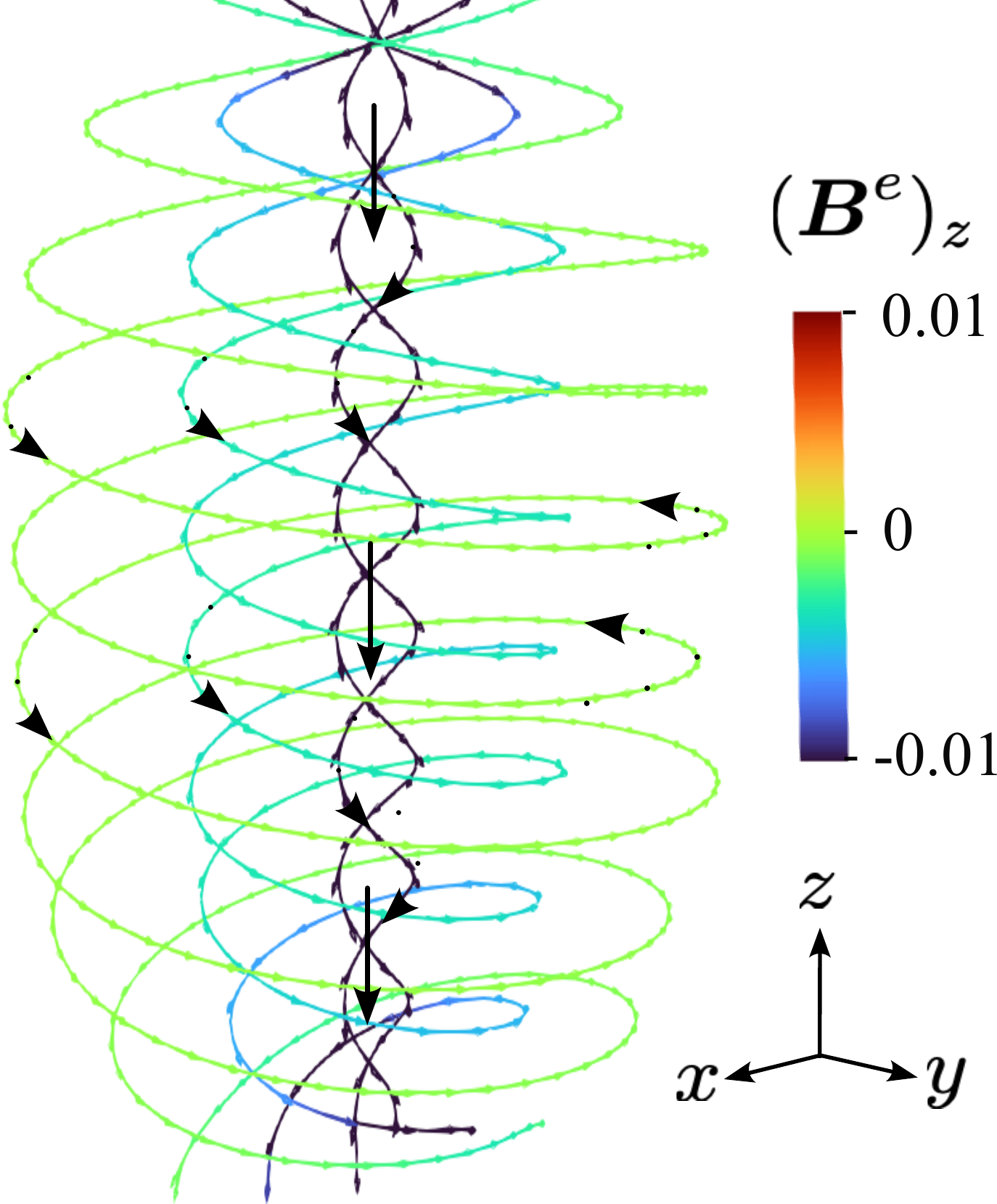}
\caption{Streamlines of the emergent magnetic field $\boldsymbol{B}^{e}$ of the stabilized TSkT displayed in the Fig. \ref{fig1} . The curves are described by circular helices with constant pitch given by $\kappa$. }
\label{fig:Bem}
\end{figure}

\end{document}